\documentclass{article}
\usepackage{graphicx}
\usepackage{amssymb,latexsym}
\usepackage{amsmath}
\begin{document}

\title{Neutrino Oscillations with Nil Mass}

\author{\ Edward R Floyd\ \\
10 Jamaica Village Road, Coronado, CA 92118-3208, USA \\
floyd@mailaps.org}

\date{9 September 2016}

\maketitle

\begin{abstract}
An alternative neutrino oscillation process is presented as a counterexample for which the neutrino may have nil mass consistent with the standard model.  The process is developed in a quantum trajectories representation of quantum mechanics, which has a Hamilton-Jacobi foundation.  This process has no need for mass differences between mass eigenstates.  Flavor oscillations and $\nu,\bar{\nu}$ oscillations are examined.

\end{abstract}

\small

\noindent PhySH: Entanglement manipulation, Neutrino oscillation, Majorana neutrino, Mass, Ab initio calculation

\medskip


\normalsize

\bigskip

\section{Introduction}

In the standard model, the neutrino, $\nu$, has nil mass since right-handed neutrinos do not exist in that model [\ref{bib:ms}].  On the other hand, the Pontecorvo, Maki, Nakagawa, and Sakata (PMNS) theory of neutrino oscillation requires mass differences among the participating neutrino's mass eigenstates whose superpositions determine the particular neutrino flavors [{\ref{bib:pont}--\ref{bib:mns}].  The mass difference of the mass eigenstates induces them to travel at different rates causing their superpositions to evolve resulting in flavor oscillation. This implies that in the neutrino only one particular mass eigenstate, at most, may have nil mass.  It follows under PMNS that all other mass eigenstates must have distinct finite masses.  While PMNS theory is sufficient to explain flavor oscillation, is it necessary?  Its explanation is extra to the standard model.  Experiments have not eliminated the possibility of neutrinos with nil mass.  We herein present an ab initio mass-neutral counterexample, for which mass may be nil, showing how nil-mass neutrinos may oscillate.

The counterexample is set in the quantum trajectories representation of quantum mechanics couched in a quantum Hamilton-Jacobi (HJ) formulation [\ref{bib:prd26}--\ref{bib:fp37a}]. Quantum trajectories, given by Jacobi's theorem, have provided insight into entanglement phenomena including nonlocal quantum trajectories of dichromatic particles and of photons in the near field of a quantum Young's diffraction experiment [\ref{bib:fp37a},\ref{bib:fp37b}].  This investigation considers that the neutrino's wave function, $\psi$, may have a dichromatic spectrum that would induce a nonlocal propagation with periodicity.  A monochromatic spectrum would manifest linear (rectilinear) motion for the neutrino that would preempt entanglement. The quantum reduced action (Hamilton's quantum characteristic function) for a neutrino with a dichromatic spectrum is explicitly shown to contain entanglement information between the two spectral components. A quantum reduced action with entanglement information permits a quantum trajectory to be nonlocal.  The nonlocality warps the quantum trajectory into alternating segments of temporally forward motion and temporally retrograde motion.  Such warping is typical of quantum trajectories of dichromatic particles [\ref{bib:fp37a}].  A retrograde segment represents a Majorana antineutrino, $\bar{\nu}$, traveling in a temporally forward direction in a Stueckelberg manner [\ref{bib:stueckelberg}].  The transitions of the quantum trajectory between the interlaced forward and retrograde segments take place at local temporal extrema where $\nu,\bar{\nu}$-pair creations or annihilations occur.  These local temporal extrema facilitate $\nu,\bar{\nu}$-oscillations. The quantum trajectory for the dichromatic neutrino is shown to exhibit  $\nu,\bar{\nu}$-oscillations independent of superposing mass eigenstate.  The quantum reduced action for a dichromatic neutrino is shown to evolve on individual segments of the quantum trajectory in a manner that permits charged current interactions to create charged leptons of different flavors at different positions along the segment.  This manifests flavor oscillations that do not need any mass differences between its PMNS mass eigenvalues.  Furthermore, flavor oscillations support a nil mass for neutrinos of any flavor. Again, nil mass neutrinos are consistent with the standard model.

Section 2 develops the formulation for relativistic quantum trajectories of dichromatic neutrinos.  The subsequent quantum trajectories give insight into $\nu,\bar{\nu}$-oscillations. The neutrino that has $\nu,\bar{\nu}$-oscillations is shown to be associated with a dichromatic $\psi$.  Entanglement between the two spectral components of the neutrino is shown to be the key.  We investigate in \S3 the quantum trajectory for an entangled 2.5 GeV neutrino with finite mass that has near luminal propagation.  We examine the $\nu,\bar{\nu}$-oscillations. In \S4 the flavor oscillations are examined.  Section 5 presents the findings, conclusions, and discussions.  Appendix A presents a brief outline for a corresponding $\psi$ representation of flavor oscillation for massless neutrinos.

\section{Formulation}

Matone [\ref{bib:m},\ref{bib:m2}] and Faraggi [\ref{bib:f}] have shown that the applicable relativistic quantum stationary Hamilton-Jacobi equation (RQSHJE)
for the neutrinos has the same form as the classical quantum stationary Hamilton-Jacobi equation (CQSHJE).  The RQSHJE for the relativistic case may be given in one-dimension $q$ for a neutrino of mass $m$ by [\ref{bib:f},\ref{bib:ijtp27}]

\begin{equation}
\left(\frac{\partial W}{\partial q}\right)^2 + m^2c^2 - \frac{E^2}{c^2} + \frac{\hbar^2}{2}\langle W;q \rangle = 0
\label{eq:rqshje}
\end{equation}

\noindent where $W$ is the relativistic quantum reduced action (Hamilton's relativistic quantum characteristic function), which is a generator of the quantum motion.  And $\langle W;q \rangle$ is the Schwarzian derivative of $W$ with respect to $q$ and is given by

\[
\langle W;q \rangle = \frac{\partial^3 W/ \partial q^3}{(\partial W/\partial q)^3} - \frac{3}{2}\left(\frac{\partial^2 W / \partial q^2}{\partial W/\partial q}\right)^2.
\]

\noindent  The Schwarzian derivative contains the quantum effects and turns the RQSHJE into a third-order nonlinear differential equation that requires
more constants of the quantum motion than the relativistic classical stationary HJ equation, which is a first-order nonlinear differential
equation [\ref{bib:prd29}].

A general solution for $W$ of Eq.\ (\ref{eq:rqshje}) is given within an integration constant by
[\ref{bib:prd34},\ref{bib:rc},\ref{bib:fm},\ref{bib:fp37a},\ref{bib:m}--\ref{bib:eh}]

\begin{equation}
W(q) = \hbar \arctan\left(\frac{A \theta(q) + B \hat{\theta}(q)}{C \theta(q) + D \hat{\theta}(q)}\right)
\label{eq:gw}
\end{equation}

\noindent where $\{\theta,\hat{\theta}\}$ is a set of independent, real solutions of the associated one-dimensional stationary Klein-Gordon equation (SKGE) for neutrinos given by [\ref{bib:f}]

\begin{equation}
-\hbar^2 c^2 \frac{\partial^2 \psi}{\partial q^2} + (m^2c^4 - E^2) \psi = 0.
\label{eq:skge}
\end{equation}

\noindent  Equations (\ref{eq:rqshje}) and (\ref{eq:skge}) remain well posed should $m=0$ in agreement with the standard model.  The coefficients $\{A,B,C,D\}$ are constants that here can be restricted to real constants and are normalized by [\ref{bib:eh}]

\begin{equation}
AD-BC=1.
\label{eq:abcdn}
\end{equation}

\noindent The Wronskian ${\mathcal W}$ is normalized so that

\begin{equation}
{\mathcal W}^2(\theta,\hat{\theta}) = (\hbar c)^{-2}.
\label{eq:wn}
\end{equation}

\noindent The normalization, $(\hbar c)^{-2}$, of the relativistic Wronskian, Eq.\ (\ref{eq:wn}), replaces the analogous normalization, $2m/\hbar^2$, of the nonrelativistic Wronskian [\ref{bib:prd34},\ref{bib:hm},\ref{bib:milne}].  (The quantum trajectory representation of quantum mechanics uses a Wronskian normalization while the $\psi$-representation uses a Born probability normalization.)   The coefficients $\{A,B,C,D\}$ are specified by its normalization, Eq.\ (\ref{eq:abcdn}),  and the initial values for quantum reduced action, $\{W,\partial_qW,\partial_q^2W\}|_{q=q_0}$ at an initial point $q_0$  analogous to those for the CQSHJE, [\ref{bib:prd29}--\ref{bib:pl450},\ref{bib:fm}].
Un-normalized coefficients also must obey $AD-BC \ne 0$, otherwise by the principle of superposition $A \theta(q) + B \hat{\theta}(q)$
and $C \theta(q) + D \hat{\theta}(q)$ would be redundant and consequently $W(q)$ would be a constant which is forbidden [\ref{bib:pl450},\ref{bib:fm}].
The general solutions of the RQSHJE and the SKGE imply each other as has been shown elsewhere for the analogous CQSHJE and
Schr\"{o}dinger equation [\ref{bib:pl450},\ref{bib:fpl9}].  One can always find a set $\{\vartheta,\hat{\vartheta}\}$ of independent solutions
for the SKGE for which the coefficients $B,C=0$ by setting $\vartheta=A\theta+B\hat{\theta}$ and $\hat{\vartheta}=C\theta+D\hat{\theta}$.  Thus,
the quantum reduced action, Eq.\ (\ref{eq:gw}), may alternatively be expressed as $W = \hbar \arctan[\vartheta(q)/\hat{\vartheta}(q)]$ by the principle of superposition for the linear SKGE.

As the conjugate momentum, $\partial_q W$, is not explicitly in the RQSHJE,  Eq.\ (\ref{eq:rqshje}), it is also a solution of the RQSHJE.  The general
form of the conjugate momentum is represented consistent with Eqs.\ (\ref{eq:rqshje}) and (\ref{eq:abcdn}) by

\begin{equation}
\frac{\partial W}{\partial q} =  \frac{\hbar (\overbrace{AD-BC}^1)\, \mathcal{W}(\theta,\hat{\theta})}{\underbrace{(A^2+C^2)\theta^2 + 2(AB+CD)\theta \hat{\theta} +(B^2+D^2)\hat{\theta}^2}_{(A\theta + B\hat{\theta})^2 + (C\theta + D\hat{\theta})^2 > 0}}
                              =  \frac{\hbar \mathcal{W}(\theta,\hat{\theta})}{(A^2+C^2)\theta^2 + 2(AB+CD)\theta \hat{\theta} +(B^2+D^2)\hat{\theta}^2}.
\label{eq:gcm}
\end{equation}

\noindent  As the denominator in Eq.\ (\ref{eq:gcm}) is always positive, as $\partial_q W$ always has the same sign as $\mathcal{W}(\theta,\hat{\theta})$, and as $\partial_q W$ is never nil, $\partial_q W$ is consistent with $W(q)$ being a monotonic function [\ref{bib:fm}].

The equation of relativistic, quantum motion is derived by Jacobi's theorem (an HJ transformation equation that provides the HJ constant coordinate $\tau$ conjugate to HJ constant momentum $E$) and is given by

\begin{equation}
\overbrace{t-\tau = \frac{\partial W}{\partial E}}^{\mbox{\scriptsize Jacobi's theorem}} = \frac{\hbar (\overbrace{AD-BC}^1)\, [\overbrace{(\partial_E \theta) \hat{\theta} - (\partial_E \hat{\theta})
\phi}^{\mathcal{W}_E(\theta,\hat{\theta})}]}{(A^2+C^2)\theta^2 + 2(AB+CD)\theta \hat{\theta} +(B^2+D^2)\hat{\theta}^2}
\label{eq:genjacobitheorem}
\end{equation}

\noindent where $t$ is time, $\tau$ is the constant coordinate that sets the epoch, $E$, and $\mathcal{W}_E$ is analogous to the Wronskian where differentiation is with respect to $E$ instead of $q$.  Equation (\ref{eq:genjacobitheorem}) also renders the one-dimensional quantum trajectory.  Analogous to non-relativistic quantum mechanics [\ref{bib:prd26},\ref{bib:fp37a},\ref{bib:fp37b}], the relativistic quantum motion generated by Eq.\ (\ref{eq:genjacobitheorem}) may be nonlocal (quantum mechanics is nonlocal).

The particular form for $W$ and $\partial W/\partial q$, using Eqs.\ (\ref{eq:gw}) and (\ref{eq:gcm}), can develop an eigenfunction, $\psi$, of the associated SKGE given by [\ref{bib:fp37a}]

\begin{equation}
\psi  = \frac{\exp(iW/\hbar)}{(\partial_q W)^{1/2}} = \left( \frac{(A^2+C^2)\theta^2 + 2(AB+CD)\theta \hat{\theta} +(B^2+D^2)\hat{\theta}^2}{ \mathcal{W}(\theta,\hat{\theta})} \right) ^{1/2} \exp \left[ i \arctan\left(\frac{A \theta + B \hat{\theta}}{C \theta + D \hat{\theta}}\right) \right].
\label{eq:gpsi}
\end{equation}

For the neutrino of energy $E$, the set of solutions necessary and sufficient to solve the SKGE is given by $\{\theta,\hat{\theta}\} = \{\sin(kq),\cos(kq)\}$ where  $k = (E^2 - m^2c^4)^{1/2}/(\hbar c)$ (note that $k$ remains well posed should $m=0$).  The generator of the motion for the neutrino is the quantum reduced action where Eq.\ (\ref{eq:gw}) may be re-expressed for neutrinos for the general solution of the RQSHJE as

\begin{equation}
W   =  \hbar \arctan{ \left( \frac{A \sin(kq) + B \cos(kq)}{C \sin(kq) + D \cos(kq)}\right)}
\label{eq:nw}
\end{equation}

\noindent or alternatively by [\ref{bib:arxiv2015}]

\begin{equation}
\sin(W) [C \sin(kq) + D \cos(kq)] = \cos(W) [A \sin(kq) + B \cos(kq)].
\label{eq:alt.nw}
\end{equation}

The general spectral resolution of the neutrino wave function $\psi$ may be found by reversing Bohm's algorithm for and only for complex wave functions [\ref{bib:bohm}], where $W=\hbar \arctan\{\Im[\psi]/\Re[\psi]\}$ and where $W$  is well posed (i.e., $AD-BC \ne 0$).  The neutrino, whose $\psi$ is given by Eq.\ (\ref{eq:gpsi}), has a dichromatic spectrum described by

\begin{equation}
\psi\  \propto \  \underbrace{\left[ \left( \frac{A+D}{2}\right) + i \left(  \frac{B-C}{2} \right) \right]}_{\mbox{\scriptsize spectral coefficient}\ a_{+k}} \exp(+ikq)\ +\  \underbrace{\left[ \left( \frac{D-A}{2}\right) + i \left( \frac{B+C}{2} \right) \right]}_{\mbox{\scriptsize spectral coefficient}\ a_{-k}} \exp(-ikq).
\label{eq:genspectrum}
\end{equation}

\noindent The spectral coefficients of Eq.\ (\ref{eq:genspectrum}) are normalized by $AD-BC=1$, Eq.\ (\ref{eq:abcdn}), consistent with the RQSHJE. For completeness, the relationships between the different sets of coefficients is given from Eq.\ (\ref{eq:genspectrum}) by

\begin{equation}
\left(
\begin{array}{cccc}
 1/2 & 1/2 &  0  &  0  \\
 -1/2& 1/2 &  0  &  0  \\
  0  &  0  & 1/2 &-1/2 \\
  0  &  0  & 1/2 & 1/2
\end{array}
\right)
\left(
\begin{array}{c}
A \\
D \\
B \\
C
\end{array}
\right) =
\left(
\begin{array}{c}
\Re[a_{+k}] \\
\Re[a_{-k}] \\
\Im[a_{+k}] \\
\Im[a_{-k}]
\end{array}
\right)
\label{eq:matrix1}
\end{equation}

\noindent and its inverse by

\begin{equation}
\left(
\begin{array}{cccc}
 1 & -1 &  0  &  0  \\
 1 & 1  &  0  &  0  \\
 0 & 0  &  1  &  1  \\
 0 & 0  &  -1 &  1
\end{array}
\right)
\left(
\begin{array}{c}
\Re[a_{+k}] \\
\Re[a_{-k}] \\
\Im[a_{+k}] \\
\Im[a_{-k}]
\end{array}
\right) =
\left(
\begin{array}{c}
A \\
D \\
B \\
C
\end{array}
\right).
\label{eq:matrix2}
\end{equation}

The set of initial values $\{W,\partial_qW,\partial_q^2W\}|_{q=q_0}$ along with $E$ are the constants of the motion that describe a unique quantum reduced action, $W(q)$.  The redundancy in the sets $\{A,B,C,D\}$ or $\{\Re[a_{+k}],\Im[a_{+k}],\Re[a_{-k}],\\
\Im[a_{-k}] \}$ is removed by the normalization Eq.\ (\ref{eq:abcdn})  with, if needed, Eq.\ (\ref{eq:matrix2}).  The sets $\{A,B,C,D\}$ or $\{\Re[a_{+k}],\Im[a_{+k}],\Re[a_{-k}],\Im[a_{-k}] \}$ are an equivalent set of constants of quantum motion that specify a unique solution of the RQSHJE while concurrently solving the auxiliary normalization Eq.\ (\ref{eq:abcdn}).

The dichromatic spectral components are coherently entangled within the neutrino's quantum reduced action.  The neutrino's generator of quantum motion must consider the neutrino's entire spectrum as represented by Eq.\ (\ref{eq:nw}). The dichromatic components $a_{+k}\exp(+ikq)$ and $a_{-k}\exp(-ikq)$ in Eq.\ (\ref{eq:matrix2}) explicitly  are not individually associated with any particular neutrino mass eigenstate nor any particular $a_{+k}\exp(+ikq)$ flavor. Rather, the spectral components $a_{+k}\exp(+ikq)$ and $a_{-k}\exp(-ikq)$ represent respectively the would-be incident wave and would-be reflected wave due to a weak force interaction if the components were unentangled.   The spectral resolution represents the entangled neutrino with flavor oscillation.  For completeness, this dichromatic $\psi$ is reminiscent of a bipolar {\itshape ansatz} of quantum trajectories [\ref{bib:prd26},\ref{bib:fm},\ref{bib:poirier}].

In general, the neutrino's quantum reduced action does not render linear motion in time [\ref{bib:fp37a}].  We briefly discuss how it can generate linear motion using the spectral coefficients of Eq.\ (\ref{eq:genspectrum}).  The conjugate momentum, $\partial_q W$, for the neutrino may be evaluated by substituting $\{\theta,\hat{\theta}\} = \{\exp(+ikq),\exp(-ikq)\}$ of Eq.\ (\ref{eq:genspectrum}) into Eq.\ (\ref{eq:genjacobitheorem}).  Likewise, with the aid of various trigonometric identities and the normalization for coefficients, Eq.\ (\ref{eq:abcdn}), $\partial_q W$ for the neutrino may be expressed in general by

\begin{equation}
\frac{\partial W}{\partial q} = \frac{\hbar k}{\frac{1}{2}(A^2+B^2+C^2+D^2) + \frac{1}{2}(A^2-B^2+C^2-D^2)\cos(2kq) + (AB+CD)\sin(2kq)},
\label{eq:dwdq}
\end{equation}

\noindent which also is another solution for to the RQSHJE as $W$ does not explicitly appear in the RQSHJE.  For $A=D=1$ and $B=C=0$, then $\psi$ is monochromatic with Eqs.\ (\ref{eq:genspectrum}) and (\ref{eq:dwdq}) implying linear motion.  The denominator on the right side of Eq.\ (\ref{eq:dwdq}) is again never nil consistent with $W$ being monotonic.

Let $\varphi$ be a phase shift.  There is a class of sets of initial values, e.g.\ $\{W,\partial_qW,\partial_q^2W\}|_{q=0} = \{\hbar \varphi,\hbar k,0\}$ for which the corresponding coefficients are given by trigonometric identities  $A,D=\cos(\varphi)$ and $B=-C=\sin(\varphi)$ [\ref{bib:dwight4012}]. The consequent neutrino's quantum reduced action would be linear, $W=\hbar kq + \hbar \varphi$.  Its wave function would be a monochromatic wave function, $\psi \propto \exp(ikq+i\varphi)$ representing a linear propagation with phase shift $\varphi$.

\section{Neutrino-Antineutrino Oscillations}

We consider a neutrino's $\psi$ that has a dichromatic spectrum whose spectral coefficients are given by $a_{+k}=\alpha$ and $a_{-k}=\beta \exp(i\phi)$ where $\alpha$ and $\beta$ are real, positive amplitudes of the spectral coefficients.  We arbitrarily choose $\alpha > \beta$ so that $W$ would monotonically increase with increasing $q$.  Had we chosen $\beta > \alpha$, then $W$ would monotonically decrease with increasing $q$. Choosing the values $\alpha=\beta$ renders a $W$ that is not well posed because then  $AD-BC=0$ and Bohm's algorithm [\ref{bib:bohm}] would not be applicable (for bound states where $\psi$ is real, see Refs.\ \ref{bib:prd34}, \ref{bib:fm}, \ref{bib:fpl9}, or \ref{bib:poirier}.  The two spectral components are entangled within the quantum reduced action for the neutrino.  This quantum reduced action is given by Eqs.\ (\ref{eq:matrix1}) and (\ref{eq:matrix2}) as

\begin{eqnarray}
W\ & = & \ \hbar\ \arctan \left(\frac{\overbrace{[\alpha-\beta \cos(\phi)]}^A\ \sin(kq)\ +\ \overbrace{\beta \sin(\phi)}^B\ \cos(kq)}{\underbrace{\beta \sin(\phi)}_C\ \sin(kq)\ +\ \underbrace{[\alpha+\beta \cos(\phi)]}_D\ \cos(kq)} \right) \label{eq:bichromaticW1} \\
   & = & \ \hbar\ \arctan \left(\frac{\alpha\ \sin(kq)\ -\ \beta\ \sin(kq-\phi)}{\alpha\ \cos(kq)\ +\ \beta\ \cos(kq-\phi)}\right) \label{eq:bichromaticW2} \\
W|_{\phi=0}   & \ne & \ \underbrace{+\ \hbar kq_{a_{+k}}}_{W_{a_{+k}}}\ \underbrace{-\ \hbar kq_{a_{-k}}}_{W_{a_{-k}}}
\label{eq:bichromaticW3}
\end{eqnarray}

\noindent The normalization of coefficients of Eqs\ (\ref{eq:bichromaticW1}) and (\ref{eq:bichromaticW2}) using Eq.\ (\ref{eq:abcdn}), is $AD-BC=\alpha^2-\beta^2=1$.  Note that $B=C$ in Eq.\ (\ref{eq:bichromaticW2}) resolves the redundancy of the coefficients, and a unique quantum reduced action is specified by $W(E,A,B,D;q)$ or $W(E,\Re[A_{+k}],\Re[a_{-k}],\Im[a_{-k}];q)$. If the two spectral components had not been entangled but represented two independent anyons and even if $\phi=0$, then by Eq.\ (\ref{eq:bichromaticW3}) the sum of their independent quantum reduced actions would still not equal the quantum reduced action of the neutrino with identical dichromatic spectrum. The dichromatic quantum reduced action inherently contains the entanglement information between the two spectral components. The inherent  entanglement of a dichromatic quantum reduced action as the generator of motion for a neutrino facilitates the possibility of nonlocal motion [\ref{bib:fp37a}].  Had there been no entanglement, then $W_{a_{+k}}$ and $W_{a_{-k}}$ of Eq.\ (\ref{eq:bichromaticW3}) would respectively represent the latent rectilinear incident and latent rectilinear reflected quantum reduced actions.

The conjugate momentum, Eq.\ (\ref{eq:gcm}), is given for the neutrino with a dichromatic spectrum by [\ref{bib:prd34},\ref{bib:fp37a}]

\begin{equation}
\frac{\partial W}{\partial q} = \frac{\hbar k}{\alpha^2 + \beta^2 + 2\alpha\beta \cos(2kq-\phi)}.
\label{eq:neutrinodwdq}
\end{equation}

\noindent  Note that the denominator on the right side of Eq.\ (\ref{eq:neutrinodwdq}) is the law of cosines if $\phi=0,\pm 2\pi,\pm 4\pi,\cdots$.  It follows that  $\partial W/\partial q > 0$, which is consistent with $W$ never being a constant [\ref{bib:pl450},\ref{bib:fm}].  Also, the conjugate momentum for the dichromatic neutrino is not a constant, as would be so for linear motion, for the denominator on the right side of Eq.\ (\ref{eq:dwdq}) is modulated by the term $2\alpha\beta \cos(2kq-\phi)$.  This variable conjugate momentum with modulation is due to the entanglement between the two spectral components [\ref{bib:fp37a}].

As developed elsewhere by a nonrelativistic analogy [\ref{bib:fp37a}], this particular form for $W$ and $\partial W/\partial q$ may be linked with an eigenfunction of the associated SKGE that is a Wronskian-normalized dichromatic wave function, $\psi$.  The solution, $\psi$, to the SKGE has compound modulation and may be expressed in alternative form [\ref{bib:prd34}] by

\begin{equation}
\psi  =  \overbrace{ \underbrace{ k^{-1/2}[\alpha^2 + \beta^2 + 2\alpha \beta \cos(2kq-\phi)]^{1/2}}_{(\partial_qW/\hbar)^{-1/2},\ \mbox{\scriptsize amplitude modulation [\ref{bib:prd34}]}}\exp \left[ i\ \underbrace{\arctan{ \left( \frac{\alpha \sin(kq) - \beta \sin(kq - \phi)}{\alpha \cos(kq) + \beta \cos(kq - \phi)}\right)}}_{W/\hbar,\ \mbox{\scriptsize wave-length modulation [\ref{bib:prd34}]}} \right]}^{\mbox{eigenfunction of SKGE}} \label{eq:npsi}.
\end{equation}

\noindent  The amplitude and wave-length modulations are not independent of each other.  The neutrino's wave function, Eq.\ (\ref{eq:npsi}), inherently contains the entanglement information of its quantum reduced action and of its conjugate momentum.  If the dichromatic spectrum represented two unentangled anyons, then their wave functions would be represented by

\[
\psi_{+k} =  k^{-1/2} \alpha \exp(+ikq_{+k})\ \ \ \mbox{and}\ \ \  \psi_{-k} = k^{-1/2} \beta \exp(i\phi) \exp(-ikq_{-k}).
\]

\noindent  The pair of independent anyons would then fly apart as one anyon, $\psi_{+k}$, would then move in an increasing $q$ direction and whose position would be denoted by $q_{+k}$ ; then the other, $\psi_{-k}$, in a decreasing $q$ direction with position $q_{-k}$.  Their combined wave functions for un-entanglement would then be $\psi_{+k} \times \psi_{-k}$.  But $\psi(q) \ne  \psi_{+q}(q_{+q}) \times \psi_{-q}(q_{-k})$ is consistent with $\psi(q)$ containing entanglement information, Eq.\ (\ref{eq:npsi}).  This is the wave function analogue of Eq.\ (\ref{eq:bichromaticW3}) for the quantum HJ representation.  The anyons (spectral components) individually represent neither any particular neutrino mass eigenstate nor any particular neutrino flavor.

Jacobi's theorem is used to determine the relativistic equation of quantum
motion [\ref{bib:prd34},\ref{bib:rc},\ref{bib:fm},\ref{bib:m}--\ref{bib:f}]. As
Jacobi's theorem also determines the equation of motion in classical mechanics,
the applicability of Jacobi's theorem transcends the division between classical and
quantum mechanics to give a universal equation of motion [\ref{bib:ijmpa15}]. Jacobi's theorem parameterizes time in the relativistic equation of quantum motion as

\begin{equation}
\underbrace{t-\tau_{\mbox{\scriptsize epoch}} = \partial W/\partial E}_{\mbox{Jacobi's theorem}} =\underbrace{\overbrace{\frac{q}{c}}^{t_{c}}\,\times\,\overbrace{\frac{E}
{(E^2-m^2c^4)^{1/2}}}^{\mbox{relativistic factor},\ H_R}}_{\small t|_{\beta=0}}\,\times\,\underbrace{\frac{1}{\alpha^2 + \beta^2 + 2 \alpha \beta
\cos(2kq-\phi)}}_{\mbox{quantum factor},\ H_Q}
\label{eq:eom}
\end{equation}

\noindent where $\tau_{\mbox{\scriptsize epoch}} $ is the HJ constant coordinate (a nontrivial
constant of integration) that specifies the epoch and where $t_{c}$ is the time needed to transit the distance $q$ by light in a vacuum.
Let us herein set $\tau_{\mbox{\scriptsize epoch}}=0$.

The relativistic factor $H_R$ is a function of $E$ and $m$, but not of the variable $q$.  The relativistic factor for  GeV neutrinos may be approximated by Eq.\ (\ref{eq:eom}) as

\begin{equation}
H_R = 1 + \frac{m^2c^4}{2E^2} + \mathcal{O}(m^4c^8/E^4) \ge 1
\label{eq:hr}
\end{equation}

\noindent where $(m^4c^8/E^4) \ll 1$. For GeV neutrinos, it is expected that $|1 - H_R| < 10^{-16}$ even for the greatest mass eigenstate of PMNS theory. In the standard model [\ref{bib:ms}], the neutrino mass is nil rendering $\lim_{m \to 0} H_ R = 1$.

The entanglement between the spectral components, Eq.\ (\ref{eq:genspectrum}), is manifested in the quantum trajectory by the cosine term in the denominator of Eq.\ (\ref{eq:eom}) analogous $\partial_qW$, Eq. (\ref{eq:dwdq}).  This entanglement remains in Eq.\ (\ref{eq:eom}) no matter how large $q$ becomes. The quantum factor, $H_Q(q)$, has extrema at

\begin{equation}
\frac{d(H_Q)}{dq} =  \frac{4k  \alpha\beta \sin(2kq-\phi)}{[{\alpha^2 + \beta^2 + 2 \alpha \beta
\cos(2kq-\phi)}]^2} = 0.
\label{eq:dfQdq}
\end{equation}

\noindent or

\begin{equation}
q = \frac{n\pi}{2k} + \frac{\phi}{2k},\ \ \ n=0,\pm1,\pm2,\cdots.
\label{eq:qn}
\end{equation}

\noindent  The entanglement information embedded in $H_Q$  induces the quantum trajectories to swing between local maximum and minimum
times of propagation [\ref{bib:fp37a}].  The quantum factor, $H_Q(q)$, has maxima given by

\begin{equation}
H_{Q,{\mbox{\scriptsize max}}} = (\alpha - \beta)^{-2}, \ \ \ q_{\mbox{\scriptsize max}} = \frac{n\pi}{k} + \frac{\phi}{2k}, \ \ \ n=0,\pm1,\pm2,\pm3,\cdots,
\label{eq:maxhq}
\end{equation}

\noindent and minima given by

\begin{equation}
H_{Q,{\mbox{\scriptsize min}}}= (\alpha + \beta)^{-2}, \ \ \ q_{\mbox{\scriptsize min}} = \frac{(2n+1)\pi}{2k} + \frac{\phi}{2k}, \ \ \ n=0,\pm1,\pm2,\pm3,\cdots.
\label{eq:minhq}
\end{equation}

\noindent  Note that $H_{Q,{\mbox{\scriptsize max}}}$ and $H_{Q,{\mbox{\scriptsize min}}}$ are functions of only the constant of the quantum motion $\beta$ for $\alpha = +(1+\beta^2)^{1/2}$.  The extrema, $H_{Q,{\mbox{\scriptsize max}}}$ and $H_{Q,{\mbox{\scriptsize min}}}$, describe an open wedge in the $q,t$-plane with its apex at the origin. The upper boundary of the wedge is given by

\[
t_{\mbox{\scriptsize upper}} = \frac{q}{c} \times H_R \times (\alpha - \beta)^{-2}.
\]

\noindent The lower boundary is given by

\[
t_{\mbox{\scriptsize lower}} = \frac{q}{c} \times H_R \times (\alpha + \beta)^{-2}.
\]

The wedge boundaries are asymptotes for the envelope (two caustics, upper and lower) of all quantum trajectories with fixed $E$ and $\beta$ but with variable phase shifts $-\pi < \phi \le +\pi$.  This wedge is denoted as the $H_Q$ wedge. The $H_Q$ wedge is interior to the envelope of all quantum trajectories.  As $q$ increases, the quantum trajectory alternates its osculations with its two bounding caustics.

The average for the quantum factor $H_Q$ of Eq.\ (\ref{eq:eom}) for a given subset of constants of the quantum motion $\{E,\beta,\phi\}$  is given by averaging over all quantum trajectories, each specified by its third constant of the motion $\phi$, that may pass through  a given $q$.  The average over $\phi$ over its range $(-\pi,+\pi)$  is given by [\ref{bib:dwight}]

\begin{eqnarray}
<H_Q(E,\beta,\phi;q)>_{\phi} & = & \frac{1}{2\pi} \int^{+\pi}_{-\pi} H_Q\,d\phi = \frac{1}{2\pi} \int^{+\pi}_{-\pi} \frac{d\phi}{\alpha^2 + \beta^2 + 2 \alpha \beta
\cos(2kq-\phi)} \nonumber \\
                             & = & \frac{1}{[(\alpha^2 + \beta^2)^2 - 4 \alpha^2 \beta^2]^{1/2}} = 1,
\label{eq:avhq}
\end{eqnarray}

\noindent normalized again by $\alpha^2 - \beta^2 = 1$.  Uniform $\phi$-averaging over all quantum trajectories that may intercept $q$ leads to an $<H_Q>_{\phi} = 1$ even if $H_{Q,{\mbox{\scriptsize max}}} > 1$.  This result is consistent with other neutrino propagation results [\ref{bib:m},\ref{bib:f},\ref{bib:11124779}] confirming that the average $H_Q$ is consistent with the expected luminal propagation.  The finding $<H_Q>_{\phi}=1$ of Eq.\ (\ref{eq:avhq}) is independent of $q$.   Note that the summation (integration) over the set of quantum trajectories represented by Eq.\ (\ref{eq:avhq}) differs with Feynman summation for the propagator.  Feynman summation does not sum over quantum trajectories.

For any caustic point $(q_{\mbox{\scriptsize caustic}},t_{\mbox{\scriptsize caustic}})$, there exists only one quantum trajectory specified by a phase shift  $-\pi < \phi_{\mbox{\scriptsize min}} \le +\pi$ that osculates with the caustic at that caustic point.  At any other points $(q_{\mbox{\scriptsize interior}},t_{\mbox{\scriptsize interior}})$ interior to the bounding envelope, there exists two quantum trajectories described by different phase shifts that cross at that interior point.  Thus, $\phi$, as a constant of quantum motion, determines the spatial placement of ripples in $W(q)$. In turn, the range of phase shifts, $-\pi < \phi \le \pi$ generates a temporal spread among the possible quantum trajectories at any $q > 0$.

\begin{figure}
\centering
\includegraphics[scale=0.75]{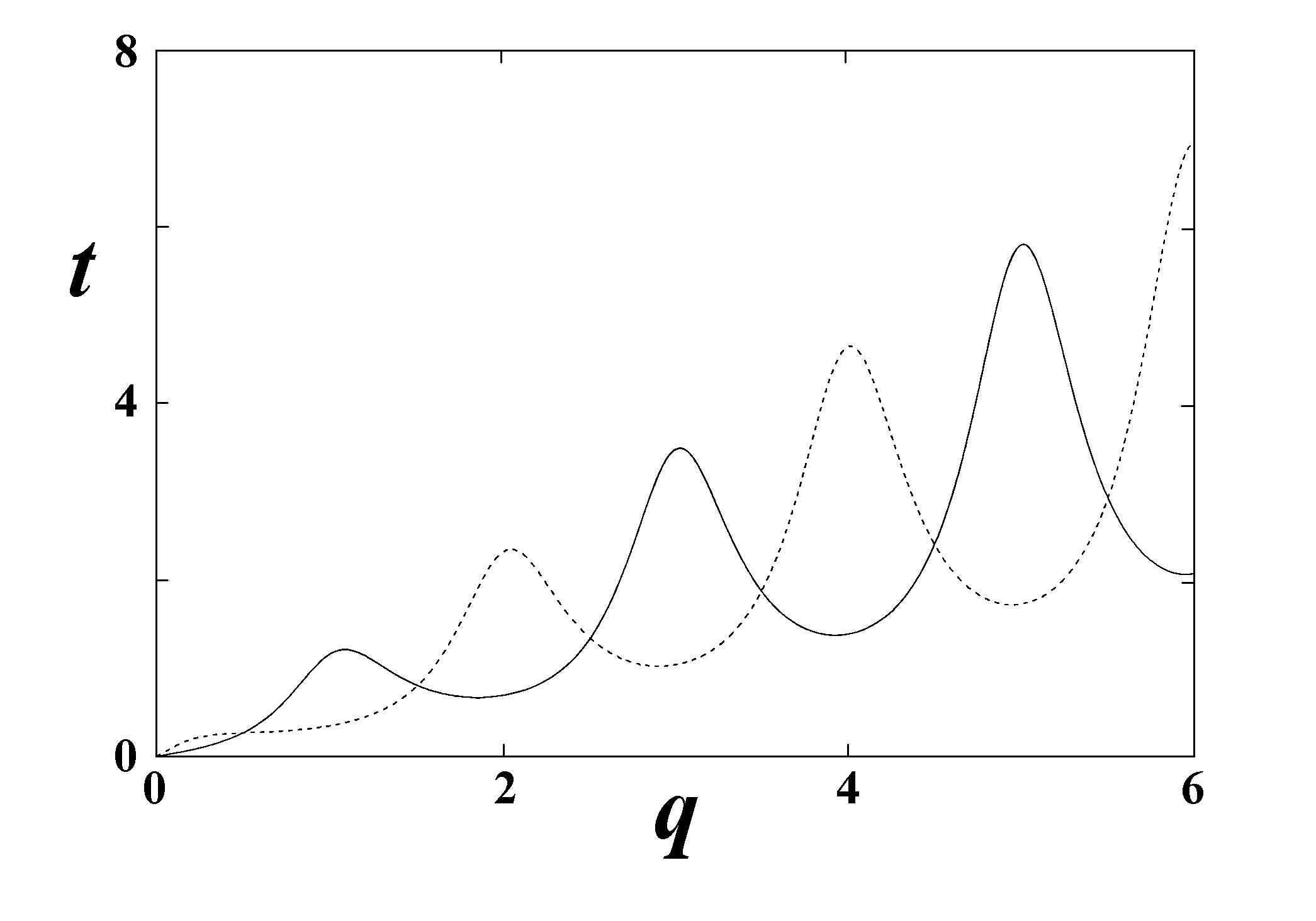}
\caption{\small Heuristic quantum trajectories are exhibited in natural units ($\hbar,c,m=1$) and for $k=\pi/2$, which fixes the constant of the quantum motion $E=(\pi^2/4+1)^{1/2}$.  The other constants of the quantum motion are given by $\beta = 0.28$, and $\phi=0,\pi$.  The quantum trajectory for $\phi=0$ is exhibited by a full line, while the quantum trajectory for $\phi=\pi$
is exhibited by a broken line.}
\end{figure}

Figure 1 exhibits the quantum trajectories for an example where $\hbar,c,m=1$ (natural units) and where the constants of the quantum motion are given
by $E=(\pi^2/4+1)^{1/2}$ (or alternatively by $E$'s proxy $k=\pi/2$), $\beta = 0.28$, and $\phi=0,\pi$ for natural units.  While this $\beta$ is many orders of magnitude too large to be representative of GeV neutrinos, it does render a good heuristic example of quantum trajectories for dichromatic particles.  The quantum
trajectories exhibit temporal turning points where the quantum trajectories alternately become temporally retrograde and forward. The entanglement
between the two spectral components, Eq.\ (\ref{eq:matrix2}), induces these transitions between temporally forward and retrograde motion.  In turn, retrograde motion after  a threshold time, $t_{\mbox{\scriptsize th}}$, may induce nonlocality as exhibited on Fig.\ 1  where the quantum trajectory permits multiple particle positions, $q$s, for selected times, $t$s.  The threshold time, $t_{\mbox{\scriptsize th}}$, for a trajectory is the first local minimum time of the trajectory where it first transitions from retrograde to forward motion.   As $q$ increases, the turning points asymptotically approach the bounds prescribed for $H_Q$, Eqs.\ (\ref{eq:maxhq}) and (\ref{eq:minhq}).  The entwined quantum trajectories for $\phi=0,\pi$ on Fig.\ 1 demonstrate the existence of trajectories with the same $\{E,\beta\}$ but with different $\phi$s that intersect a particular $q$ at different transit times, albeit trajectories also cross at interior $q,t$ points.  For completeness, these quasi-periodic meanderings on Fig.\ 1 asymptotically approach periodicity as $q \to \infty$ [\ref{bib:fp37a}].

While early latent, temporally retrograde segments may be suppressed as shown by Fig. 1, temporally retrograde segments are realized as $q$ increases.  Any finite $\beta$ will induce nonlocality for sufficiently large $q$ [\ref{bib:fp37a}]. This may be shown by investigating the equation of quantum motion, Eq.\ (\ref{eq:eom}). Temporal turning points are smooth and exist where the reciprocal velocity, $dt/dq$, goes to zero. The behavior of the reciprocal velocity may be derived
from Eq.\ (\ref{eq:eom}) and given by

\begin{equation}
\frac{dt}{dq} = \frac{d[q(c^{-1}H_R)]}{dq} \times H_Q\, + \, qc^{-1}H_R \times \frac{d(H_Q)}{dq}.
\label{eq:dtdq}
\end{equation}

\noindent The conditions for a temporal turning point, $dt/dq = 0$, may be simplified to

\begin{equation}
\frac{1}{q}\ +\ \frac{4\alpha \beta k \sin(2kq-\phi)}{\alpha^2 + \beta^2 + 2\alpha \beta \cos(2kq-\phi)} = 0.
\label{eq:sdtdq}
\end{equation}

\noindent As $q$ may increase without bound in Eqs.\ (\ref{eq:dtdq}) and (\ref{eq:sdtdq}), the $1/q$ term in Eq.\ (\ref{eq:sdtdq}) will
decrease sufficiently to ensure the existence of temporal turning points for any finite $\beta$. The existence of temporally retrograde motion follows.  For completeness, the conjugate momentum, Eq.\ (\ref{eq:neutrinodwdq}), obeys $\partial W/\partial q > 0,\ \alpha > \beta > 0$, even when the quantum trajectory transits a local temporal extremum.  Hence, the quantum reduced action is never a constant but always increases with $q$ even on retrograde segments [\ref{bib:pl450},\ref{bib:fm}].  The displacement in $q$ between temporal turning point and its associated extremum of $H_Q$, Eq.\ (\ref{eq:dfQdq}), which is given by the $q^{-1}$ term in Eq.\ (\ref{eq:sdtdq}), gets less with increasing $q$.  Still, the the quantum trajectory in the $q,t$ plane firsts osculates with the caustic, next passes through the extrema in $H_Q$, and then reverses its temporal direction at the temporal tuning point.  These three points asymptotically converge.

The set of constants of the quantum motion $\{E,\beta,\phi\}$ specifies the quantum trajectory, Eq.\ (\ref{eq:eom}). The energy $E$ is an input into the relativistic factor $H_R$ and through its proxy, $k$, determines the wavelength of the meanders of the quantum factor $H_Q$ by the distance, $\pi/(2k)$, between successive maxima, Eq.\ (\ref{eq:maxhq}), or successive minima, Eq.\ (\ref{eq:minhq}).  The apex angle of the open $H_Q$ wedge is a function of $\beta$.  Consequently, $\beta$ is an input to the duration of retrograde segments, if existing, of the quantum trajectory, Eqs.\ (\ref{eq:maxhq}) and (\ref{eq:minhq}).  The phase, $\phi$, determines the placement of the temporal retrograde segments within the quantum trajectory, Eqs.\ (\ref{eq:maxhq}) and (\ref{eq:minhq}).  The two quantum trajectories exhibited on Fig.\ 1 for example have phase shifts that differ by $\pi$, which interchanges where their forward and retrograde segments occur.  As the two trajectories are out of phase by $\pi$, their local temporal maxima and minima turning points are nearly co-located in $q$ but not $t$. As $q \to \infty$ this approximate co-location asymptotically approaches exact by Eqs.\ (\ref{eq:maxhq}) and (\ref{eq:minhq}).

There is an alternative Stueckelberg interpretation of segments of temporal retrograde motion [\ref{bib:stueckelberg}]: such retrograde motion is the forward motion of an antiparticle, here an antineutrino, $\bar{\nu}$, moving forward in time.  Under this interpretation, $\nu,\bar{\nu}$-pair creations occur at the temporal
turning points (local temporal minimum) associated with $H_{Q,{\mbox{\scriptsize min}}}$ while $\nu,\bar{\nu}$-pair annihilations
occur at the temporal turning points (local temporal maximum) associated with $H_{Q,\mbox{\scriptsize max}}$.

The Majorana neutrino, $\nu$ is its own antineutrino, $\bar{\nu}$.  The Majorana neutrino is a neutral left-handed fermion that as such interacts as a neutrino.  On the other hand, Majorana antineutrino also has the same mass and one-half spin but is right-handed and so only interacts as an antineutrino.  Under Majorana theory, it is only the handedness, left or right, that determines whether it be neutrino or antineutrino.  All neutrinos are left-handed; antineutrinos, right-handed.  All else (mass and half-spin) are the same for the neutrino and the antineutrino.  The PMNS theory would require the neutrino and antineutrino to have different masses for a $\nu,\bar{\nu}$-oscillation to exist.  In contrast, the quantum trajectory, as exhibited on Fig.\ 1, conserves mass by iso-mass $\nu,\bar{\nu}$-oscillations at temporal turning points. This conservation of mass over the neutrino's quantum trajectory completes making the neutrino a Majorana fermion. The quantum trajectory of the Majorana neutrino as exhibited on Fig.\ 1 is continuous with interlaced $\nu$-segments and $\bar{\nu}$-segments.  Thus, a Majorana neutrino may oscillate between acting as a Majorana neutrino and acting as a Majorana antineutrino without changing mass.  Furthermore, the neutrino mass may be nil consistent with the standard model.

For similar exposition of neutrino behavior, Horwicz and Aharonovich [\ref{bib:ha}] used Stueckelberg's covariant relativistic theory [\ref{bib:lrt}] to develop neutrino trajectories with retrograde segments due to relativistic rather than quantum causes.   Horwicz and Aharonovich describe the interlacing of forward and retrograde trajectory segments to be neutrino oscillations.

Also the quantum trajectories of massless photons, in the quantum Young's experiment with internal interference between the contributions of the double slits, have been shown to exhibit temporal retrograde segments interspersed between forward temporal segments in the near field [\ref{bib:fp37b}].  Hence, neutrinos under a quantum trajectory representation need not have a finite mass to exhibit oscillations.  Again, if the neutrino mass is nil, then $H_R=1$ in the equation of quantum motion, Eqs.\  (\ref{eq:eom}) and  (\ref{eq:hr}).

The restrictions of Cohen and Glashow [\ref{bib:cg}] are inapplicable herein.  In the vicinity of the temporal turning points where $\dot{q}$ becomes instantaneously infinite, Cohen and Glashow predict a loss of energy due to Cherenkov-like $e^-,e^+$-pair creation.  The quantum trajectory is continuous through the temporal turning point where  $\nu,\bar{\nu}$-pair creations or annihilations occur instead of Cohen and Glashow $e^-,e^+$-pair creations.  Cohen and Glashow had considered $\nu,\bar{\nu}$-pair creations in the form $\nu_{\mu} \rightarrow \nu_{\mu}+\nu_e+\bar{\nu}_2$, but dismissed them reporting that neutrinos of all flavors propagate ``at virtually the same value" [\ref{bib:cg}].  For quantum trajectories, Faraggi and Matone developed the effective mass $m_Q$ given by [\ref{bib:fm}]

\[
m_Q \equiv \frac{\partial_q W}{\dot{x}} = m\left( 1-\frac{\hbar^2}{2} \frac{\partial \langle W;q \rangle}{\partial E} \right).
\]

\noindent  The conjugate momentum, Eq.\ (\ref{eq:gcm}) is always finite definite.  For forward temporal motion, $m_Q$ is positive definite; for retrograde motion, $m_Q$ is negative definite; and at temporal extrema, $m_Q=0$ [\ref{bib:fp37a}].  Therefore, pair creation at local temporal minima would not be an endoergic process. And pair annihilation at local temporal maxima would not be an exoergic process [\ref{bib:fp37a}].

\section{Example}

Let us examine the quantum trajectory of a 2.5 GeV neutrino to illustrate how oscillations develop for high-energy neutrinos with small $\beta$.  We arbitrarily set $\beta = 5 \times 10^{-7}$ and $\phi = 0$ in Eq.\ (\ref{eq:bichromaticW1}).  The smallness of $\beta$ generates a very narrow $H_Q$ wedge, Eqs.\ (\ref{eq:eom}), (\ref{eq:maxhq}) and (\ref{eq:minhq}).  With $\phi=0$, then the coefficients are given as

\[
A = \left( \frac{\alpha-\beta}{\alpha+\beta}\right)^{1/2}, \ \ B,C=0, \ \ \mbox{and} \ \ D = \left( \frac{\alpha+\beta}{\alpha-\beta}\right)^{1/2},
\]

\noindent and the spectral coefficients are given by Eq.\ (\ref{eq:matrix1}) as

\[
a_{+k} = (A+D)/2 = \alpha \ \ \mbox{and} \ \  a_{-k} =(D-A)/2 = \beta.
\]

\noindent With these coefficients, the quantum reduced action simplifies to

\begin{equation}
W = \hbar \arctan\left( \frac{\alpha-\beta}{\alpha+\beta} \tan(kq) \right).
\label{eq:xw}
\end{equation}

\noindent Even with a small $\beta = 5 \times 10^{-7}$, $a_{-k}$ is still finite. This is sufficient for eventual nonlocal propagation as shown in this
example's development.

The neutrino mass for any flavor for a 2.5 GeV neutrino has a negligible effect upon its quantum trajectory, smaller than $10^{-18}$.  For 2.5 GeV and a neutrino rest mass of 2 eV, the neutrino would have a wavelength given by $\lambda= hc (E^2-m^2c^4)^{-1/2} \approx 495$ am; for nil rest mass, $\lambda \approx 495$ am also.   While the quantum trajectory for a simulated neutrino with rest mass 2 eV and constants of the quantum motion $E=2.5$ GeV, $\beta = 5 \times 10^{-7}$ and $\phi=0$, is investigated herein, the results for nil mass neutrinos would differ relatively by only $10^{-18}$.  Segments of its computed quantum trajectory are presented in Figs. 2(a), 2(b), and 2(c).

\begin{figure}
\centering
\includegraphics[scale=0.30]{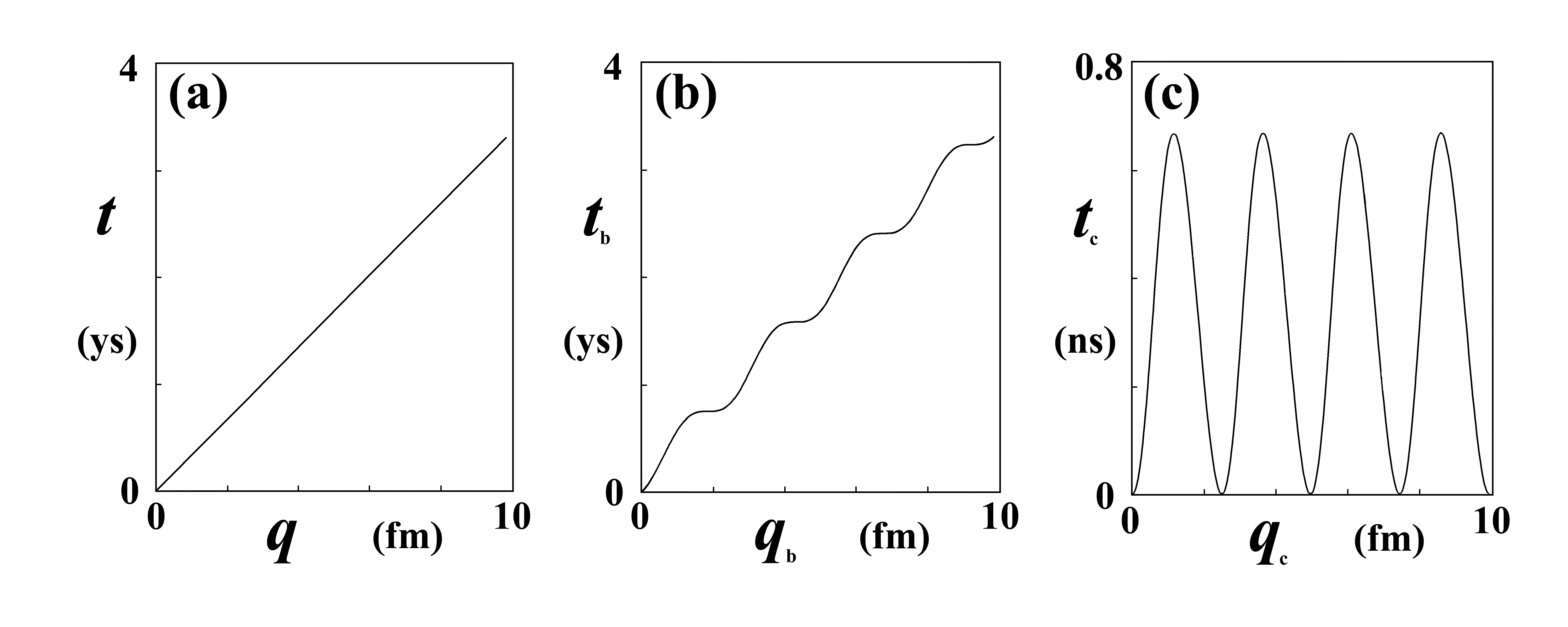}
\caption{\small The initial 2 wavelengths ($\lambda$) are exhibited on 2(a) and show that there is insufficient entanglement due to a finite $\beta$
too small to develop temporal retrograde segments in the first 2 $\lambda$.  On 2(b) is a 2 $\lambda$ section of a theoretical quantum
trajectory that simulates the region where the transition to temporal
retrograde motion begins. The displayed origin of this section for 2(b) is at 80\,000 wavelengths, ($q_{\mbox{\scriptsize b,origin}} = 39.67$ pm and
$t_{\mbox{\scriptsize b,origin}} = 13.23$ zs). Hence, the coordinates for 2(b) are $q_{\mbox{\scriptsize b}} =q-q_{\mbox{\scriptsize b,origin}}$ and
$t_{\mbox{\scriptsize b}}=t-t_{\mbox{\scriptsize b,origin}}$.  On 2(c) is a 2 $\lambda$
section of a theoretical quantum trajectory with sufficient entanglement to simulate temporal retrograde motion and neutrino oscillation. The displayed origin of this section for 2(c) is at 100 km.
Hence, the coordinates for 2(c) are $q_{\mbox{\scriptsize c}}=q-100$ km and $t_{\mbox{\scriptsize c}}=t-333.6 \mu$s.
The time scale for 2(c) differs with those of 2(a) and 2(b).}
\end{figure}

Figure 2(a) shows the quantum trajectory for the first 2 wavelengths. To the eye, it appears to be a straight line, but it does have some microscopic
quasi-periodic meanders within its bounding envelope approximated by the narrow $H_Q$ wedge.  These meanders are insufficient
to cause time reversals in Fig.\ 2(a).  Figure 2(b) exhibits the quantum trajectory for 2 wavelengths
beginning 80\,000 wavelengths ($q = 39.67$  pm) from its origin at $q=0$ and in the transition region where time reversals are about to begin to appear.
The quasi-periodic meanderings of the quantum trajectory are apparent in Fig.\ 2(b).  The time coordinate (vertical axis) of Fig.\ 2(b) is $t_{\mbox{\scriptsize b}}=t-68.93$ zs while the distance coordinate (horizontal axis) is $q_{\mbox{\scriptsize b}}=q-39.67$ pm.  Figure 2(c) shows the quantum trajectory
at 100 km from its origin. Retrograde motion has now appeared.  Segments of the quantum trajectory alternate between forward and retrograde motion with regard to $q$.  Between segments, there exist smooth turning points near the upper and lower caustics of the $H_Q$ wedge where neutrino speed becomes instantaneously infinite for $dt/dq \to 0$ at local temporal extrema points consistent with Eqs.\ (\ref{eq:dtdq}) and (\ref{eq:sdtdq}).  Nevertheless, the neutrino speed remains integrable as
substantiated by the existence of $t$, Eq.\ (\ref{eq:eom}), as exhibited by  Fig.\ 2(c).  The coordinates for 2(c) are $q_{\mbox{\scriptsize c}}=q-100$ km and $t_{\mbox{\scriptsize c}}=t-333.6 \mu$s.   Note that the time scale of Fig.\ 2(c) is much coarser, by a degree of $10^{15}$, than those of Figs.\ 2(a) and 2(b).  This accommodates the size of the durations of the temporally forward and retrograde segments.  While not apparent in Fig.\ 2(c), corresponding points on successive wavelengths advance in time about 1.654 ys consistent with Figs.\ 2(a) and 2(b).

Note in Fig.\ 2(c) that the temporal duration of the retrograde                                                                                                                                                                                                                                                                                                                                                                                                                                                                                                                                                                                                                                                                                                                                                                                       segments has increased to about 0.6671 ns. These durations are sufficiently large so that if $0 \le t_{\mbox{\scriptsize c}} \le 0.6671$ ns, then the projection of $t_{\mbox{\scriptsize c}}$ upon the quantum trajectory intercepts it at multiple values of $q_{\mbox{\scriptsize c}}$, which exhibits nonlocality. Also, these durations will continue to increase with increasing $q$ as the quantum trajectory proceeds out the open $H_Q$ wedge.  The segments of the quantum trajectories between
$\nu,\bar{\nu}$-pair creations in the vicinity of $H_{Q,\mbox{\scriptsize min}}$ and $\nu,\bar{\nu}$-pair annihilations in the
vicinity of $H_{Q,\mbox{\scriptsize max}}$ are spatially approximately 16.23 am long and have an approximate temporal duration of 0.6671 ns as previously noted.

For $\beta \ne 0$, Figs.\ 2(b) and 2(c) each exhibit that $W$ has 4 periodic meanderings in 2 $\lambda$ segment (neighboring meanderings are nearly identical).  This spatial frequency doubling is manifested by the doubled wave number, $2k$, in $\partial_q W$ that is observed in Eq.\ (\ref{eq:neutrinodwdq}) for $\beta \ne 0$ ($\partial_q W$ is also a solution of the nonlinear third-order RQSHJE).  Concurrently, the wave number for $\psi$ of the associated linear SKGE, Eq.\ (\ref{eq:skge}), is just $k$.  Should $\beta = 0$ in Eq.\ (\ref{eq:xw}) for rectilinear propagation, the corresponding RQSHJE would then have been reduced to first-order for its higher-order term $\langle W;q \rangle$ would have been nulled out.

\section{Flavor Oscillation}

Neutrinos are only subject to the weak force, which is very short ranged, and to gravity.  The neutrino travels unimpeded through vacuum and matter
until it undergoes a weak force interaction, either a rare charged current interaction or the even rarer neutral current interaction.
The neutrino is not observed directly but detected only by its interactions with matter.  A charged current interaction requires
that the antineutrino have sufficient energy ($>1.8$ MeV) to create its corresponding charged lepton.  The charged lepton, either electron, muon, or tau,
created by a charged current interaction identifies the flavor of the neutrino, either $\nu_e,\ \nu_{\mu}$ , or $\nu_{\tau}$ respectively.
The form of the charged current interaction is dependent upon the particular charged lepton created.  So, a change of outcome of a
charged current interaction for a neutrino as a function of its position $q$ along its quantum trajectory can manifest an oscillation
between neutrino flavors in the quantum trajectory representation.  Linear propagation, $\beta=0$ in Eq.\ (\ref{eq:xw}), would preempt flavor oscillation.

While $\nu,\bar{\nu}$-oscillations occurred at temporal extrema as discussed in \S3 and \S4, flavor oscillations are dependent upon the evolution of
$\{W(q),\partial_qW(q),\partial^2_qW(q)\}$ with $q$.  The neutrino's quantum reduced action for any current or neutral interaction must maintain $\mathcal{C}^2$
continuity over the range of any interaction to produce a post-interaction quantum reduced action of the partner charged lepton, for the RQSHJE is a third order differential equation.  A classical precedent of maintaining $\mathcal{C}^2$ continuity of the quantum reduced action while tunneling has already been presented [\ref{bib:afldb20}].  Such interactions must also account for the quantum motion of the target charged lepton.   The Mikheyev-Smirnov-Wolfenstein (MSW) effect [\ref{bib:msw1},\ref{bib:msw2}] may increase the interaction cross-section in dense matter.

Let us consider an inverse beta decay ($\bar{\nu}_e + p \rightarrow n + e^+$) instigated by the 2.5 GeV neutrino whose trajectory was described by Fig.\ 2(c). The 2.5 GeV neutrino acts as an antineutrino in a Stueckelberg manner on retrograde segments of the quantum trajectory exhibited on Fig.\ 2(c) and as a neutrino on the forward segments [\ref{bib:stueckelberg}].  The antineutrino under Stueckelberg propagates forward in time on the neutrino's retrograde segment of the quantum trajectory, cf.\ Fig 2(c).  The antineutrino must assume a $\bar{\nu}_e$ flavor while approaching the point $q_p$ on the ``retrograde" segment to be able to initiate and consummate an inverse beta decay upon a proton located at $q_p$. The amplitude, $\beta$ of the $a_{-k}$ spectral coefficient is such to pre-empt any latent ``reflection" at the initiation of the inverse beta decay.  The entanglement between the dichromatic spectral components fully compensates for any would-be reflective reaction from the inverse beta decay.  If $\beta$ is sized so that $\beta \exp(-ikq)$ is a matching substitute for the would be reflected wave, then the interaction product is only the inverse beta decay. The antineutrino's quantum reduced action must maintain $\mathcal{C}^2$ continuity throughout  cosummating the inverse beta decay. The values of $\{W(q),\partial_qW(q),\partial^2_qW(q)\}$ at the start of the inverse beta decay must also be consistent with the quantum motion of the target proton.  A neutrino propagating nonlinearly offers an evolving set of $\{W(q),\partial_qW(q),\partial^2_qW(q)\}$ that determines the flavor selection as a function of $q$.  The quantum reduced action, $W(q)$  described by Eq.\ (\ref{eq:xw}}), for the 2.5 GeV (anti)neutrino evolves monotonically with periodic meanderings from linearity.   Consequently, the conjugate momentum, $\partial_q W(q)$, is not constant but evolves periodically as prescribed by

\begin{equation}
\frac{\partial W(q)}{\partial q} = \frac{\hbar k}{\alpha^2 + \beta^2 + 2 \alpha \beta \cos(2kq)}.
\label{eq:xcm}
\end{equation}

\noindent The conjugate momentum remains positive finite consistent with Faraggi and Matone [\ref{bib:fm}] even during periods of retrograde quantum motion exhibited on Fig.\ 2(c).  The periodic evolution of $\partial^2_qW(q)$ is given by

\begin{equation}
\frac{\partial^2 W(q)}{\partial q^2} = \frac{4 \hbar k^2 \alpha \beta \sin(2kq)}{[\alpha^2 + \beta^2 + 2 \alpha \beta \cos(2kq)]^2}.
\label{eq:xwqq}
\end{equation}

\noindent Equations (\ref{eq:neutrinodwdq}) and (\ref{eq:xw})--(\ref{eq:xwqq}) describe the evolution of $\{W(q),\partial_qW(q),\partial^2_qW(q)\}$ for a $\mathcal{C}^2$ continuous quantum reduced action.  The values of $\{W(q),\partial_qW(q),\partial^2_qW(q)\}$ evolve with a common spatial periodicity given by the doubled wave number $2k$.  The values of $\{W(q),\partial_qW(q),\partial^2_qW(q)\}$ also determine the particular flavor as a function of $q$.  The particular phase of the evolution of $\{W(q),\partial_qW(q),\partial^2_qW(q)\}$ is subject to the particular initial conditions $\{W,\partial_qW,\partial^2_qW\}|_{q=q_0}$ necessary and sufficient to solve the RQSHJE, Eq.\ (\ref{eq:rqshje}), uniquely.  In the rare event when the values of $\{W(q),\partial_qW(q),\partial^2_qW(q)\}$ are compatible with quantum motion of the target proton, inverse beta decay happens. Again, should $\beta = 0$, then linear neutrino propagation would follow and flavor oscillation would be preempted.

Flavor oscillation occurs as the set $\{W(q),\partial_qW(q),\partial^2_qW(q)\}$ evolves with $q$ among $q=q_e$, $q=q_{\mu}$, and $q=q_{\tau}$ where the (anti)neutrino assumes the properties of respectively the $\nu_e$, $\nu_{\mu}$ and $\nu_{\tau}$. The specifications of the domains $\forall q \in \{q_e\}$, $\forall q \in \{q_{\mu}\}$, and $\forall q \in \{q_{\tau}\}$ are beyond the scope of this opus.  The domains may be energy dependent and dependent upon the quantum motion of the target particle.  The domains may be segmented and may overlap.  The variables, $q_e,\ q_{\mu}$, and $q_{\tau}$ may be continuous or discrete.  The union  $\{q_e\} \cup \{q_{\mu}\} \cup \{q_{\tau}\}$  may neither span nor cover the domain of the (anti)neutrino's quantum trajectory.

\section{Findings, Conclusions, and Discussions}

\paragraph{Findings:}  By counterexample, a quantum trajectory process for a neutrino describes flavor oscillations and $\nu,\bar{\nu}$ oscillations that does not need any mass difference among the mass eigenstates of the neutrino. Nor does the process need a superposition of mass eigenstates.  While PMNS theory implies mass and is extra to the standard model, an alternative mass-neutral theory for neutrino oscillations does exist.

\paragraph{Conclusions:} Flavor oscillations independent of mass differences also implies that neutrino mass may be nil consistent with the standard model. The quantum trajectory model implies that the Majorana hypothesis for the neutrino-antineutrino pair is correct. Any finite deviation from linear propagation of the neutrino (e.g., $\beta \ne 0$) will induce flavor oscillations and, at some finite distance, $\nu,\bar{\nu}$-oscillations.

\paragraph{Discussions:} Spin did not play any direct role in the quantum trajectory description of neutrino oscillations.  A quantum Hamilton-Jacobi analysis describes  neutrino oscillations.  The Dirac equation was also not needed.  On the other hand, the PMNS description of neutrino oscillation also needed neither spin nor the Dirac equation.

The cost of the counter example, massless neutrinos, is the requirement that the neutrino be dichromatic.  A monochromatic neutrino would imply rectilinear propagation with neither flavor nor $\nu,\bar{\nu}$ oscillations.  As dichromatic neutrinos are permitted solutions in both the wave representation, Eq.\ (\ref{eq:skge}), and the quantum Hamilton-Jacobi representation, Eq.\ (\ref{eq:gw}) [\ref{bib:fp37a},\ref{bib:pe5}], the cost is only choosing the proper dichromatic solution.

This opus was developed in a quantum trajectories representation with an underlying quantum Hamilton-Jacobi foundation.  The counter example presented herein, which explains flavor oscillations with  massless neutrinos, is not an anomaly unique to the methodology of quantum trajectories.  The counter example may also be derived in principle by wave mechanics as outlined in Appendix A.

In summary, a neutrino interaction entangles its incident spectral component with its Majorana reflected spectral component that has Stueckelberg retrograde motion.  The resultant ``self-entangled" neutrino propagates nonlinearly inducing oscillations and nonlocality until it interacts with matter where its quantum reduced action can maintain $\mathcal{C}^2$ continuity throughout the interaction.

\bigskip

\small

\begin{enumerate}\itemsep -.06in

\item \label{bib:ms} Mohapatra R N and Smirnov A Y: Neutrino Mass and New Physics. {Ann.\ Rev.\ Nucl. Part.\ Sci.} {\bfseries 56}, 569--628 (2006), hep-ph/0603118.

\item \label{bib:pont} Pontecorvo B: Mesonium and Antimesonium { Sov.\ Phys.\ JETP} {\bfseries 6}, 429--31 (1958); in Russian, {Zh.\ Eksp.\ Teor.\ Fiz.} {\bfseries 33}, 549--57 (1957).

\item \label{bib:pont2} Pontecorvo B: Neutrino Experiment and the Problem of Electronic Charge.  {Sov.\ Phys.\ JETP} {\bfseries 26}, 984--8 (1968); in Russian, {Zh.\ Eksp.\ Teor.\ Fiz.} {\bfseries 53}, 1717--1725 (1967).

\item \label{bib:mns} Maki B, Nakagawa N and Sakata S: Remarks on theUnified Model of Elementary Particles. {Prog. Theor. Phys.} {\bfseries 28}, 870--80 (192).

\item \label{bib:prd26}  Floyd E R: Modified potential and Bohm's quantum potential  {Phys.\ Rev.} D {\bfseries 26}, 1339--47 (1982).

\item \label{bib:prd29} Floyd E R: Arbitrary Initial Conditions of Hidden Variables. {Phys.\ Rev.} D {\bfseries 29}, 1842--4 (1984).

\item \label{bib:prd34} Floyd E R: Closed Form Solutions for the Modified Potential. {Phys.\ Rev.} D {\bfseries 34}, 3246--9 (1986).

\item \label{bib:pl450} Faraggi A E and Matone M: Quantum mechanics from an equivalence principle. {Phys.\ Lett.} B{\bfseries 450}, 34--40 (1999),  hep-th/9705108.

\item \label{bib:rc} Carroll R: Some Remarks on Time, Uncertainty, and Spin. {J.\ Can.\ Phys.} {\bfseries 77}, 319--25 (1999), quant-ph/9903081).

\item \label{bib:fm}  Faraggi A E and Matone M: The Equivalence Postulate of Quantum Mechanics.  2000 {Int.\ J.\ Mod.\ Phys.} A {\bfseries 15}, 1869--2017 (2000), hep-th/9809127.

\item \label{bib:bfm} Bertoldi G, Faraggi A E and Matone M: Equivalence Principle, Higher Dimensional M\"{o}bius Group and
the Hidden Antisymmetric Tensor of Quantum Mechanics.  {Class.\ Quant.\ Grav.} {\bfseries 17}, 3965--4006 (2000), hep-th/9909201.

\item \label{bib:fp37a}  Floyd E R: Interference, Reduced Action, and Trajectories. 2007 {Found.\ Phys.} {\bfseries 37}, 1386--402 (2007), quant-ph/0605120v3.

\item \label{bib:fp37b}  Floyd E R: Welcher Weg? A Trajectory Representation of a Quantum Young's Experiment. {Found.\ Phys.} {\bfseries 37}, 1403--20 (2007),  quant-ph/0605121v3.

\item \label{bib:stueckelberg} Stueckelberg E C G: La signification du temps propre en mécanique ondulatoire. {Helv.\ Phys.\ Acta.} {\bfseries 14}, 51--80 (1941).

\item \label{bib:m} Matone M: Superluminal neutrinos and a curious phenomenon in the relativistic Hamilton-Jacobi equation. (2011), arXiv:1109.6631v2.

\item \label{bib:m2}  Matone M: Neutrino speed and temperature. (2011) arXiv:1111.0270v3.

\item \label{bib:f}  Faraggi A E: OPERA data and the equivalence postulate of quantum mechanics. (2011), arXiv:1110.1857v2.

\item \label{bib:ijtp27} Floyd E R: Progress in a Trajectory Interpretation of the Klein-Gordon Equation. {Int.\ J.\ Theor.\ Phys.} {\bfseries 27}, 273--81 (1988).

\item \label{bib:fpl9} Floyd E R: Where and Why the Generalized Hamilton-Jacobi Representation Describes Microstates of the Schr\"{o}dinger Wave Function. {Found.\ Phys.\ Lett.} {\bf 9}, 489--97 (1996), quant-ph/9707051.

\item \label{bib:hm}  Hecht C E and Mayer J E:: Extension of the WKB equation. {Phys.\ Rev.} {\bfseries 106}, 1156--60 (1953).

\item \label{bib:milne} Milne W E: The numerical determination of characteristic numbers. {Phys.\ Rev.} {\bfseries 35}, 863--7 (1930).

\item \label{bib:eh} Hille E: {Ordinary Differential Equations in the Complex Plain} (Dover: Mineola, NY, 1976) pp 374--401.

\item \label{bib:arxiv2015} Floyd E R: Quantization, Energy Quantization, and Time Parametrization. (2015), arXiv:1508:00291.

\item \label{bib:bohm}  Bohm, D.: A Suggested Interpretation of the Quantum Theory in Terms of ``Hidden" Variables, I. {Phys.\ Rev.} {\bfseries 85}, 166--179 (1953).

\item \label{bib:poirier} Poirier B: Reconciling semiclassical and Bohmian mechanics. I. Stationary states. {J.\ Chem.\ Phys.} {\bfseries 121}, 4501--4515 (2004).

\item \label{bib:dwight4012}  Dwight H B: {Tables of Integrals and Other Mathematical Data.} 4th ed.\ (MacMillan:
New York, 1961) \P401.2.

\item \label{bib:ijmpa15} Floyd, E R: {Classical Limits of the Trajectory Interpretation of Quantum Mechanics, Loss of Information and Residual Indeterminacy.} Int.\ J.\ Mod.\ Phys.\ A {\bfseries 15}, 1563--1568 (2000), quant-ph/9907092.

\item \label{bib:dwight} Dwight H B: 1961 {Tables of Integrals and Other Mathematical Data} 4th ed.\ (MacMillan:
New York, 1961) \P 858.520, \P 858.521, \P 858.524, and \P 858.525.

\item \label{bib:11124779} Floyd E R: OPERA Superluminal Neutrinos per Quantum Trajectories (2011), arXiv:1112.4779v2.

\item \label{bib:ha} Horwicz L P and Aharonovich I: Neutrinos and $v<c$. arXiv:1203.1632v9 (2012).

\item \label{bib:lrt} Lacki J, Ruegg H and Telegdi V: The Road to Stueckelberg's Covariant Perturbation Theory as Illustrated by Successive Treatments of Compton Scattering. {Stud.\ Hist.\ Philos.\ Mod.\ Phys.} {\bfseries 30}, 457--518 (1999) physics/9903023.

\item \label{bib:cg} Cohen A G and Glashow S L: New Constraints on Neutrino Velocities. {Phys.\ Rev.\ Lett.} {\bfseries 107}, 181803 (2011), arXiv:1109.6562.

\item \label{bib:afldb20} Floyd E R: A Trajectory Interpretation of Tunneling. {An.\ Fond.\ L.\ de Broglie} {\bfseries 20}, 263--279 (1955).

\item \label{bib:msw1} Mikheev S P and Smirnov A Yu: Resonance enhancement of oscillations in matter and solar neutrino spectroscopy. {Sov. J. Nuc. Phys.} {\bfseries 42}, 913–917 (1985).

\item \label{bib:msw2} Wolfenstein L: Neutrino oscillations in matter. {Phys. Rev.} {\bfseries D 17}, 2369 (1978).

\item \label{bib:pe5} Floyd E R: Comments on Mayant's ``A note on Bohm's Interpretation of quantum mechanics", Phys. Essays {\bfseries 5}, 130--2 (1992).

\end{enumerate}

\bigskip

\normalsize

\appendix

\section{Outline for a $\boldsymbol{\psi}$ Representation of Flavor Oscillation}

A suggested $\psi$ representation of the quantum trajectories algorithm for neutrino oscillation is now outlined.  While the quantum reduced action (a generator of quantum motion) of the quantum trajectories representation accounts for the entanglement between the dichromatic components of the neutrino's spectrum, a composite $\psi$, Eq.\ (\ref{eq:gpsi}), does the same accounting in the $\psi$ representation.  In both representations, the two spectral components, $k_{\pm} = \pm k = \pm (E^2 - m^2c^4)^{1/2}/(\hbar c)$, are not manipulated as separate entities but compositely to incorporate their mutual entanglement with each other to describe the dichromatic neutrino's behavior.  As the neutrino is not bound, its $\psi$ is complex (if bound, then $\psi$ would be real) [\ref{bib:fpl9},\ref{bib:bohm}].  Consequently, its complex $\psi$ does not have any microstates in the quantum trajectories representation [\ref{bib:fpl9}]. When encountering an interaction, the $k_{-}$ spectral component of $\psi$ is the proxy for the would-be reflected wave. In this manner, the would-be reflected wave may be considered to be a Majorana entity.  This would-be reflected wave acts as a secondary, complementary wave that is entangled with the primary would-be incident wave, which is represented by $k_{+}$ spectral component.  In other words, the would-be reflection is already incorporated into the neutrino.  The entanglement between the two spectral components produce a dichromatic wave function with compound modulation, Eqs.\ (\ref{eq:gpsi}) and (\ref{eq:genspectrum}).  In the $\psi$ representation, the neutrino propagates until it encounters a matching flavor-dependent interaction where the complex dichromatic $\psi$ and $\partial_q \psi$ are continuous, $\mathcal{C}^1$, across the interaction [the wave-length and amplitude modulations are not independent of each other, Eq.\ (\ref{eq:npsi})].  Flavor oscillations are incorporated into the dichromatic $\psi$ for the neutrino by compound modulation, Eqs.\ (\ref{eq:genspectrum}) and (\ref{eq:npsi}).  As wavelength modulation, $W(q)$, and amplitude modulation, $[\partial_qW(q)]^{-1/2}$, periodically evolve with $q$ [\ref{bib:prd34}], the values for the dichromatic $\psi(q)$, and $\partial_q W(q)$ change with $q$ to produce periodic flavor oscillation.  For a deeper development of this algorithm, the interested reader is invited to review Ref.\ \ref{bib:afldb20}, which discusses the non-relativistic tunneling problem
from the points of view of both the quantum trajectories and the $\psi$ representations.

\end{document}